# Controlling Synthesis of Nanostructured Silver Aggregates by Light

J. Brummer, R. Langlois, M. Loth, A. K. Popov, R. Schmitz, G. Taft, R. Tanke, and A. Wruck
*Departments of Chemistry, Physics & Astronomy, and Biology, University of Wisconsin-Stevens Point,*
*Stevens Point, Wisconsin 54481;*
apopov@uwsp.edu

**Abstract:** The possibilities for control over the size and properties of silver nanoaggregates with incoherent and laser light are investigated. The applications in nanoengineering and for giant enhancement of optical processes at nanoscale are discussed.

Metal nanostructured aggregates were demonstrated to enable confinement and giant enhancement of various optical processes at subwavelength scale (see, e.g., [1-4] and references therein). Such breaking of the diffraction limits has important applications in nanophotonics and for microlasers, in optical electronics and biophotonics, for sensing of molecules and for photomodification of biological objects. We have studied photostimulated synthesis of nanostructured aggregates consisting of hundreds to thousands of silver nanoparticles. The study is aimed at investigation of the possibilities for control over the size and properties of such nanoparticles and their aggregates.

In our experiments, colloidal nanoparticles were created by two methods resulting in the reduction of AgNO3 in the solution. Initial aggregation in the colloid was controlled by amount of stabilizer used during synthesis. In the first method ethanol was used for reduction and polymer PVP [(poly)vinyl pyrrolidone] for initial stabilization. In the second method silver nanoparticles were created through reduction by NaBH4 in the presence of stabilizing citrate and BSPP [Bis(p-sulfonatophenyl)phenylphosphine dihydrate dipotassium salt].

Light can stimulate further change of a size and shape of the nanoparticles [5,6] as well as can greatly increase the rate of their aggregation into nanostructured metamaterials. The entire process is determined by a number of underlying contributing processes in electrolyte, which include plasmon excitation, photoeffect, and a shielding role of polymers (for review see, e.g. [4]). The structure and properties of such objects resemble statistical fractals characterized by the fractal dimensionality. Increased aggregation rates usually leads to larger and more rarefied structures possessing smaller fractal dimensionality. However, the outcomes may strongly depend on the synthesis method. Our objective is to investigate how the way in which the aggregates cluster together depends on the source of radiation, stabilizers, and on the synthesis method. This may lead to engineering of silver and gold nanostructures, which fractal dimensionality and consequently optical properties can be manipulated by light and polymers. As outlined, this has expected direct applications in nano-sensors and nano-lasers.

Interaction between metal nanoparticles causes perturbation and shift of their plasmon resonances. The smaller is distance between the particles, the larger is shift. Owing to realization of various distances between the particles inside the aggregate, cluster formation leads to inhomogeneous broadening of the absorption resonance, which depends on distribution of distances between the nanoparticles inside the aggregate. A long tail stretched to the red side of the absorption spectrum indicates appearance of closely spaced nanoparticles. Indeed, interference process associated with the nearfields of optically excited nanoparticles leads to giant enhancement of optical process at nanoscale provided by such nanostructured objects. If necessary, they can be further fixed in soft solid matrices. Thus manipulating structures of such objects presents an important goal of optical nanoengineering.

Evolution of absorption spectrum of the colloid allows one to monitor the aggregation stages. At the same time, spectral broadening and stretching the spectrum to the red side indicates appearance of the properties that enable subwavelength concentration of local optical fields. Nonmonochromatic light may enhance aggregation rate, whereas strong laser radiation stimulate synthesis and simultaneously may cause photomodification of the synthesized aggregates [1-4]. In our experiments, aggregation was stimulated by irradiation from 70 W Mercury UV lamp and from 20 mW Argon-ion laser. The aggregation stages were monitored through the absorption spectra. The size and shape of the aggregates as well as of the constituent nanoparticles were determined directly with transmission electron microscopy. Some of the experimental results are presented at the figures below.

Figure 1a shows details of evolution of the colloid absorption spectrum caused by spontaneous aggregation in the dark (intermediate curve) and by photoinduced aggregation (most broad resonance). It is seen that pronounced absorption is developed in the far red wing of the spectrum with the aid of control ultraviolet radiation. Transmission Electron Microscopy images, presented in Figure 1b directly prove the appearance of the aggregates and opportunity for control over synthesized structures with light.

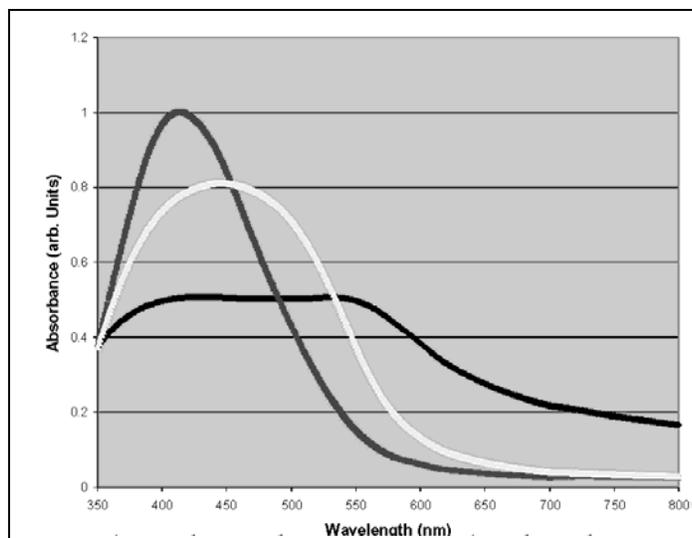

Fig. 1a. Method 1. Evolution of the absorbance spectrum of silver colloids caused by irradiation with UV lamp. Narrow peak – initial spectrum, broad – after exposure to UV lamp for 33 days, intermediate – control sample after 33 days in the dark.

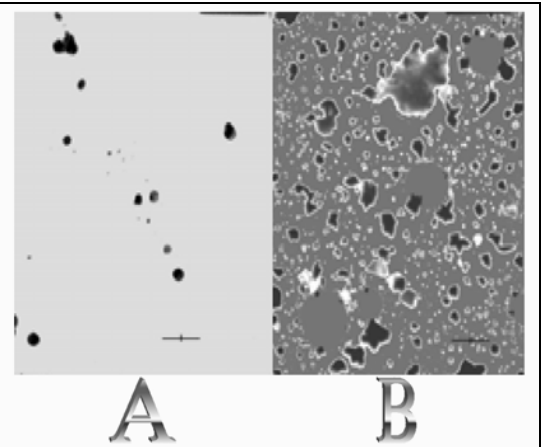

Fig. 1b. Corresponding TEM images. (A) Sample kept in the dark (bar equals 25 nm) (B) Plethora of large aggregates from sample exposed to UV light for 33days (bar equals 2 µm)

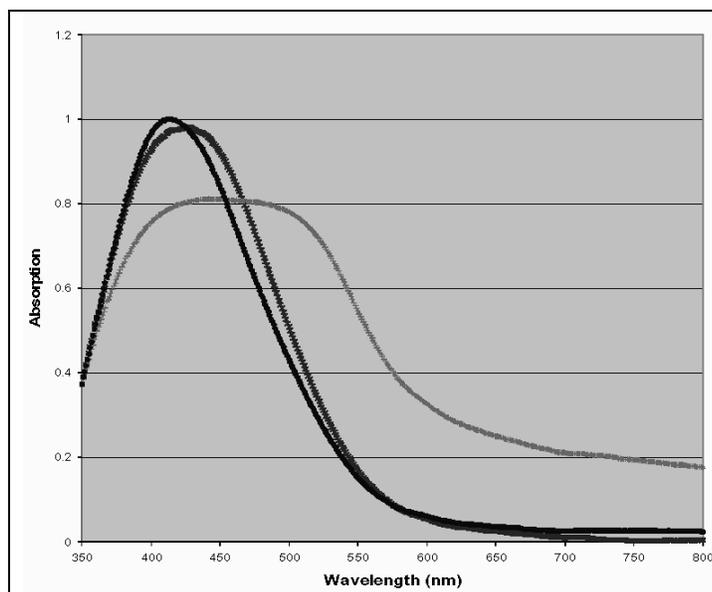

Fig. 2a. Method 1. Narrow peak – initial spectrum of the colloid, broad – after 24 hours exposure to Argon-ion laser irradiation, intermediate – control sample kept in the dark for 24 hours.

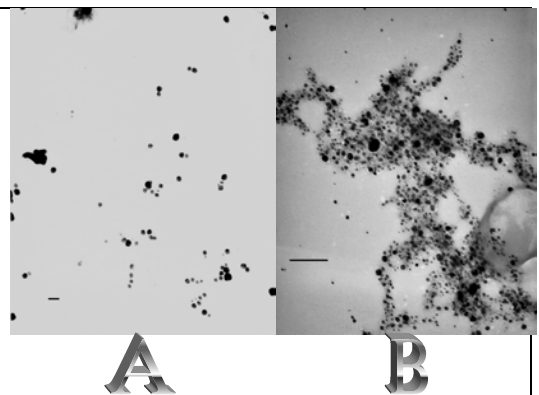

Fig. 2b. Corresponding TEM images. (A) nanoparticles kept in the dark for 24 hours (bar equals 25 nm). (B) The same after irradiation by Argon-ion laser for 24 hours (bar equals 5 µm).

Figure 2a demonstrates much faster aggregation stimulated with continuous wave Argon-ion laser, and Fig. 2b displays somewhat different properties of the synthesized structure.

The experiments have revealed substantial dependence of photostimulation and of properties of the produced aggregates on the initial properties of the colloid. It is demonstrated in Fig. 2. The possible reasons are discussed in [4], and are the subject of our further investigations.

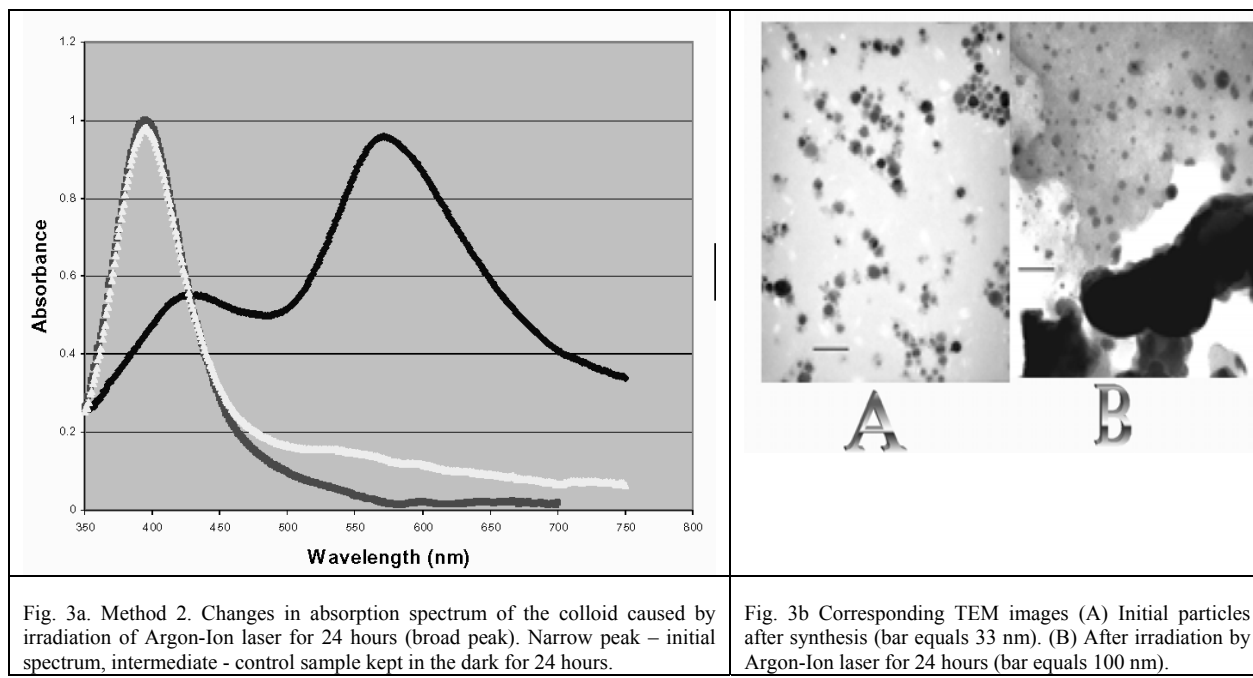

Fig. 3a. Method 2. Changes in absorption spectrum of the colloid caused by irradiation of Argon-Ion laser for 24 hours (broad peak). Narrow peak – initial spectrum, intermediate - control sample kept in the dark for 24 hours.

Fig. 3b Corresponding TEM images (A) Initial particles after synthesis (bar equals 33 nm). (B) After irradiation by Argon-Ion laser for 24 hours (bar equals 100 nm).

Thus, the displayed results confirm the feasibilities of manipulating such metal nanostructures.

The work was supported in part by the grant MDA972-03-1-0020 from DARPA through Purdue University. Electron Microscopy support for this project provided by an L&S UEI grant.

\* Undergraduate students.